\documentclass[aps,prb,twocolumn,superscriptaddress,citeautoscript]{revtex4-2}
\usepackage[dvipdfmx]{graphicx}
\usepackage{subfigmat}
\usepackage[whole]{bxcjkjatype}

\usepackage{amsmath,ams symb} 
\usepackage{color}
\usepackage{bm}
\usepackage{physics}
\usepackage{hyperref}
\hypersetup{
	citecolor=blue,
	colorlinks = true,
	urlcolor = blue
}
\usepackage[english]{babel}
\usepackage{comment}
\usepackage[normalem]{ulem}

\begin{document}
\selectlanguage{english}

\title{Parametric instability in a magnomechanical system}
\author{Takahiro Uto}
\affiliation{Institute for Quantum Electronics, ETH Z\"{u}rich, CH-8093 Z\"{u}rich, Switzerland}

\author{Daigo Oue}
\email{daigo.oue@gmail.com}
\affiliation{%
Kavli Institute for Theoretical Sciences, University of Chinese Academy of Sciences, Beijing, 100190, China.
}
\affiliation{Instituto de Telecomunica\c{c}\~{o}es, Instituto Superior T\'{e}cnico, University of Lisbon, 1049-001 Lisboa, Portugal}
\affiliation{The Blackett Laboratory, Imperial College London, London SW7 2AZ, United Kingdom}
\affiliation{RIKEN Centre for Advanced Photonics, Saitama 351-0198, Japan}

\date{\today}

\begin{abstract}
We study parametric instability in a magnomechanical system, specifically examining magnon tunneling between moving ferromagnetic insulators. Our analysis reveals that quantum fluctuations generate spin currents above a critical velocity threshold, while no spin currents occur below this threshold at low temperatures. The critical velocity depends on magnon stiffness and Zeeman energy. Approaching the threshold, the spin current becomes divergent, linked to the $PT$-symmetry-breaking transition. This enhanced behavior could offer high-sensitivity measurements and efficient spin current generation in magnon-based quantum technology.
\end{abstract}
\maketitle 

\section{introduction}
Parametric instability is a ubiquitous phenomenon in nature. One pivotal example is the hydrodynamic instability of motions of fluids\,\cite{drazin2002introduction, kumar1994parametric}. In fluid dynamics, instability refers to the phenomenon where a fluid experiences an amplification of small perturbations from its equilibrium state, disrupting the system's behavior under certain conditions. This instability in the fluid extends even to the scope of cosmology\,\cite{allahverdi2010reheating}. Another important example is in nonlinear optics. The light produced by nonlinear interactions is efficiently amplified by satisfying the photon energy and phase-matching requirements\,\cite{boyd2008nonlinear}. This phenomenon has been widely used, for example, in laser frequency conversion\,\cite{boyd2008nonlinear}, optical parametric amplifier\,\cite{boyd2008nonlinear, hansryd2002fiber}, and generation of entangled photons by spontaneous parametric down-conversion (SPDC)\,\cite{kwiat1995new}. 
Regarding signal amplification, optomechanics acts as a vital venue\,\cite{massel2011microwave, aspelmeyer2014cavity}. Two-mode squeezing interaction lies at the heart of parametric amplification. In the absence of dissipation, this induces simultaneous excitation in both mechanical and optical oscillations, leading to an exponential growth of the energies stored in the system\,\cite{metelmann2023parametric}.
However, on the other hand, in the other fields\,(i.e. electron spin system), the investigation of coherent and efficient amplification using parametric instability remains sparse.

Magnomechanics\,\cite{zhang2016cavity, lachance2019hybrid, yuan2022quantum}, as a hybrid quantum system which involves mechanical degrees of freedom\,\cite{clerk2020hybrid}, has recently attracted considerable attention due to their compatibility with electronic devices\,\cite{barman20212021, pirro2021advances, yuan2022quantum} as well as their capabilities of magnon Bose-Einstein condensation~(BEC) and spin superfluidity\,\cite{bunkov2010magnon}, magnon squeezing\,\cite{li2019squeezed}, and entanglement generation\,\cite{li2018magnon, zhang2019quantum}.
This field is closely related to and inspired by optomechanics that deals with photons instead of magnons and has offered versatile applications in quantum technology, such as quantum transducers and quantum sensing\,\cite{aspelmeyer2014cavity, barzanjeh2022optomechanics, lauk2020perspectives}.
It would be worthwhile to investigate if parametric-instability-based phenomena could be observed in such systems and would be instrumental for magnon-based quantum technology, such as quantum transducers\,\cite{bejarano2024parametric}, quantum memories\,\cite{tanji2009heralded}, high precision measurements\,\cite{wolski2020dissipation} and logic gates\,\cite{chumak2014magnon, yuan2022quantum}, through efficient amplification of relevant physical quantities of interest.

\begin{figure}
    \centering
    \includegraphics[width=\linewidth]{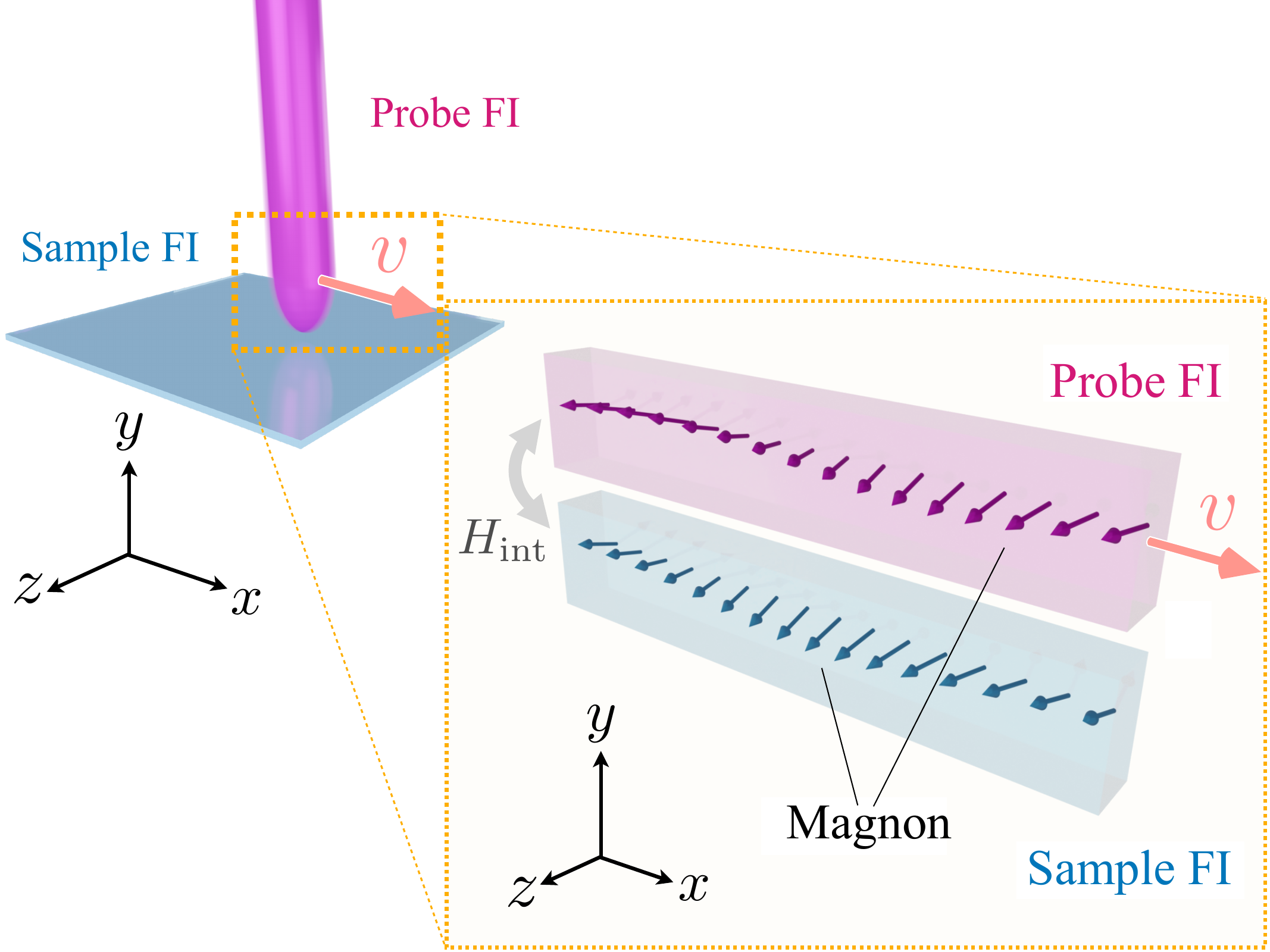}
    \caption{Schematic figure of the setup to investigate the interaction between a ferromagnetic insulator (FI) probe operating in lateral mode and an FI sample. Simplified setups are shown in the inset, where two one-dimensional FIs holding magnons are placed with a narrow gap. While the sample FI is at rest, the probe FI is moving in the $x$ direction at a constant velocity $v$. }
    \label{fig1}
\end{figure}

In this work, we theoretically investigate magnomechanical systems in terms of magnon tunneling between two ferromagnetic insulator media.
Figure\,\ref{fig1}(a) illustrates a schematic picture of a system in which a ferromagnetic microcantilever vibrates in a lateral mode near the sample.
Based on the fact that the response of the electron spin system is on the order of pico-seconds, which is fast enough compared to the time scale of the Doppler shift and slow enough for spin dynamics, we simplify this system as a model where one of two one-dimensional ferromagnetic media is sliding at a constant speed, and the other is at rest as shown in Figure\,\ref{fig1}(b).
The following discussion may be adiabatically applied to non-inertial motion\,(i.e., we can substitute $v \rightarrow v(t)$ if $\dot{v} \ll 1$); however, the generalization of our theory\,(e.g., inclusion of non-adiabatic effect) is left to future work.
 We analyze this model as a tunneling problem, taking into account that the dipolar and exchange interactions are dominant when the distance between the two media is close enough.
 We demonstrate that two magnons in different media strongly interact to drive the spin transfer if the sliding velocity exceeds a critical value which is determined by material parameters.
 Furthermore, we also discuss the analogy to the parametric instability at the critical point, where the spin current diverges.
 The combination of this phenomenon and magnon tunneling transport can be used to generate spin currents with high efficiency, leading to high-sensitivity measurements of spin information at a surface and spin current amplification.

\section{Spin non-conservative Hamiltonian.}
We consider two media separated by a narrow gap. The sample ferromagnetic insulator~(Sample FI) is at rest, while the probe one~(Probe FI) is moving at a constant speed~$v$~(Figure\,\ref{fig1}(b)). Each medium hosts a magnon and is described by the Hamiltonian using the lowest-order Holstein-Primakoff theory: $H_0 = \sum_{k\eta} \hbar\omega_{\eta, k} b_{\eta, k}^\dagger b_{\eta, k}$,
where $\eta=\mathrm{p}, \mathrm{s}$ specify the medium~($\mathrm{p}$ and $\mathrm{s}$ correspond to Probe FI and Sample FI, respectively), and $\omega_{\eta, k}$ is the magnon dispersion relation, and $b_{\eta, k}^\dagger\,(b_{\eta, k})$ creates$\,$(annihilates) magnons in the $\eta$ medium.
Note that we can safely apply the spin-wave approximation if our probe is on the order of or bigger than the typical size of the magnetic domain $\sim 1 \mathrm{\mu m}$\,\cite{kittel1946theory}.

The interaction of the two magnets is described in terms of magnons at the lowest order in the Holstein-Primakoff theory.
There are two interaction channels, which are represented by the following Hamiltonian:
\begin{align}
    H_{\rm int} &= H_1 + H_2, \\
    &= \sum_k \qty(H_{\rm ex}b_{\mathrm{s},k}^\dagger b_{\mathrm{p},k} + \rm H.c.) \nonumber \\
    &~~~+ \sum_k \qty(H_{\rm nc}b_{\mathrm{s},k}b_{\mathrm{p},-k} + \rm H.c.) \nonumber,
\end{align}
where $H_{\rm ex}$ and $H_{\rm nc}$ are coupling strengths originated from exchange and dipole-dipole interaction. 
Note that spin is conserved (not conserved) in the first (second) channel $H _ {1} (H _ 2)$.
Here, for simplicity, we consider the constant coupling strengths.
Note also that we shall adopt the lowest-order Holstein-Primakoff theory and elucidate the effect of the shearing motion in the system.

\section{Doppler effect on the spin current}
Let us consider the case of $T\rightarrow0$ and lossless limit. By utilizing the nonequilibrium (Schwinger-Keldysh) Green's function, we perturbatively evaluate the spin current flowing into the sample medium, which is defined as the temporal variation in the number of spins within the sample medium~(See Appendix \ref{appx:derivation-spin-current} for an evaluation of the spin current), $I_{\rm nc} \equiv \hbar\sum _ k \partial{(b _ {\mathrm{s},k} ^ \dagger b _ {\mathrm{s},k})}/\partial{t}$.
Up to the second order in the interaction strength $H_{\rm nc}$, we can write
\begin{align}\label{eq: current}
    \expval{I_{\rm nc}} = \frac{8\pi H_{\rm nc}^2}{\hbar}\sum_k \delta(\omega_{\mathrm{p},k} + \omega_{\mathrm{s},-k}).
\end{align}
We should note that the contribution of $H_1$ vanishes in the low-temperature ($T\rightarrow0$) limit, while the $H _ 2$ contribution remains. In this sense, we can say this spin current~\eqref{eq: current} is coming purely from the quantum fluctuation (See Appendix \ref{appx:derivation-spin-current}).

The present configuration shares similarities with the one for quantum friction\,\cite{pendry1997shearing,pendry2010quantum,milton2016reality} in that (i) both involve shearing motion between two bodies without physical contact; (ii) the Doppler effect plays a central role in driving tunnelling transport between the two bodies. However, it differs from the quantum friction scenario in that: (I) the focus is on spin transport, while the quantum friction problem involves momentum transport; (II) the carrier in our case is a magnon, whereas theirs is a photon and/or polariton, resulting in different dispersions. In the theory of quantum friction, surface plasmon polaritons ($\omega \simeq \omega _ \mathrm{sp}$) play a central role in the quasi-static (short-wavelength) limit ($k \gg 1$)\,\cite{pendry2010quantum,brevik2022fluctuational}, while the dispersion of the spin carrier (magnon) in the present case becomes constant only at the long-wavelength limit ($k \ll 1$) as considered below. 

Here, we measure the spin-current coming into the sample medium. From the view on the sample medium, the probe medium is moving at constant velocity $v$, leading to the Doppler shift of the excitation $\omega_{\mathrm{p},k} \rightarrow \omega_{\mathrm{p},k} - vk$. 
Thus, the argument of the delta function, $\Omega _ k = \omega _ {\mathrm{p},k} + \omega _ {\mathrm{s},-k}$, in Eq.\,\eqref{eq: current} should also be shifted,
$\Omega _ k \rightarrow \Omega _ k - vk$.
Considering a parabolic dispersion for the magnon $\omega_{\eta, k} = Dk^2 + \omega_0$ for simplicity, we can explicitly write the condition under which the argument of the delta function vanishes, leading to the generation of the spin current: 
\begin{align}
    0 = \omega _ {\mathrm{p},k} + \omega _ {\mathrm{s},-k} - vk = 2Dk^2 - vk + 2\omega_0.
    \label{eq:condition}
\end{align}
The influence of the Doppler shift on the system is represented by $vk$, allowing us to interpret Eq.\,\eqref{eq:condition} can be viewed as the energy conservation relation in our proposed spin current generation.
It is also worth noting that the Doppler effect solely frequency-modulates the system without imparting any momentum; consequently, the generated magnons propagate in opposite directions (we can observe that one of the magnons is labeled with $k$ while the other with $-k$), ensuring the total momentum is conserved.

The quadratic equation\,\eqref{eq:condition} has a solution if the two parabolic dispersion curves, $\omega _ {\mathrm{p},k} - vk$ and $\omega _ {\mathrm{s},-k}$, tangentially intersect.
The condition for this tangential contact reads
\begin{align}
    |v| \geq 4\sqrt{D\omega_0} \equiv v_\mathrm{cr}.
    \label{eq: threshold}
\end{align}
This inequality means that the spin current generation (creation of magnons) has a critical velocity $v_\mathrm{cr}$, and below $v_\mathrm{cr}$, there is no spin current. This discussion is reminiscent of the Landau criteria for superfluidity\,\cite{pethick2008bose}, where two fluids with different stream velocities are specified, and a critical phenomenon (superfluid to normal fluid transition) at a threshold velocity is described in terms of the competition between the two fluids. In the theory of superfluidity, Bogoliubov phonons are created above the threshold velocity, leading to a transition to the normal fluid phase\,\cite{pethick2008bose}. 

\begin{figure}[h!]
    \centering
    \includegraphics[width=\linewidth]{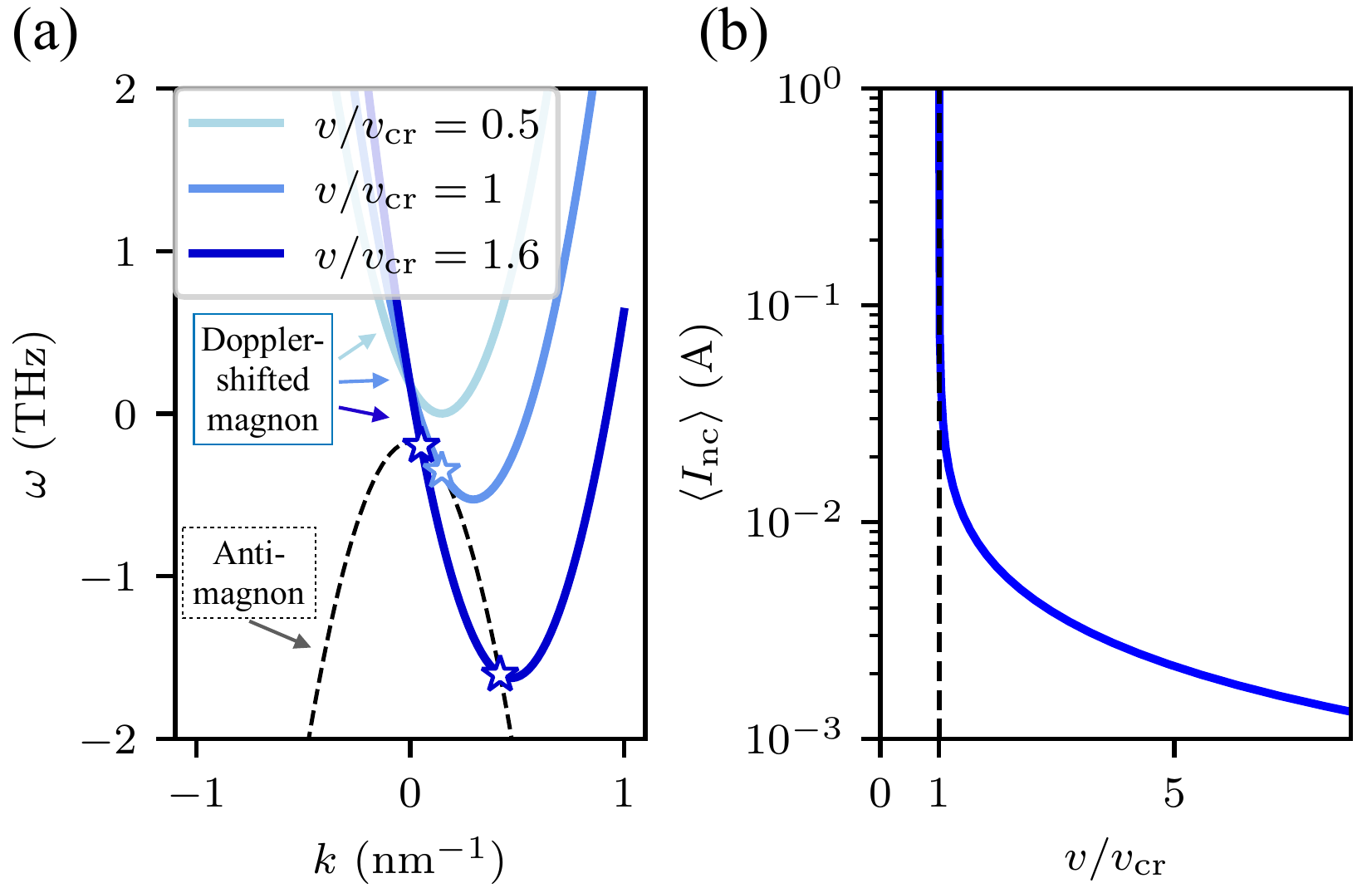}
    \caption{(a)\,Crossing of the dispersion relations of magnons in the sample (dashed line) and probe medium (solid line), which are responsible for the spin current generation, at different velocities of probe magnon $v$. At the crossing point(s), the two magnons strongly interact to drive the spin transfer. The dispersion relation of the probe magnon shifts as a function of velocity $v$ due to the Doppler effect. Therefore, if there is no crossing point depending on the value of $v~(v < v_\mathrm{cr})$, no spin transfer occurs. (b)\,Spin current as a function of the velocity of probe magnon $v$ normalized by a critical velocity $v_\mathrm{cr}$. Parametric instability at $v_\mathrm{cr}$ causes divergence of the spin current.}
    \label{fig2}
\end{figure}

Crossing of the dispersion relations of magnons in the sample and probe medium for various sliding velocities $v$ is shown in Figure\,\ref{fig2}(a). The crossing points contribute to the spin current~\eqref{eq: current}. In the laboratory frame, the sample medium is at rest, while the probe medium is moving at a constant speed; hence, the frequencies of magnons on the probe medium apparently decrease due to the Doppler effect.
As a result, the magnon frequency in the probe medium ($\omega_{\mathrm{p},k} - vk$) can be negative as in the \v{C}erenkov radiation\,\cite{silveirinha2013quantization,horsley2016negative, svidzinsky2021unruh}, where the Doppler shift plays a similar role, and photon frequency can be apparently negative.

By introducing the effective mass of the magnon $D=\hbar/2m^*$, we can rewrite the condition for the intersection\,\eqref{eq: threshold} as $m^*v^2/2 \geq 2\Delta$.
This expression provides another perspective: the `kinetic' energy provided by the sliding motion $m ^ * v ^ 2/2$ should be equal to or greater than \textit{twice} the `gap' energy, $\Delta = 2\hbar\omega _ 0$, between the lowest frequency mode in the sample medium and the highest frequency mode in the probe medium to get the spin current. This implies that our spin current is parametrically excited.
Remind that the parametric excitation necessitates twice the fundamental excitation frequency.

The spin current per unit length can be explicitly evaluated as
\begin{align}
    \expval{I_{\rm nc}}
    &= \frac{8 H_{\rm nc}^2}{\hbar}
    \sum _ k
    \sum _ {\Omega _ q = vq}
    \frac{\delta(k - q)}{\abs{\dv*{(\Omega _ k - vk)}{k}}}
    \label{eq:vHs}
    \\
    &= \frac{8 H_{\rm nc}^2}{\hbar} \frac{1}{\sqrt{v^2-v_\mathrm{cr}^2}} ~~(\mathrm{if} ~|v| \geq v_\mathrm{cr}).
\end{align}
Remind that $\Omega _ k = \omega _ {\mathrm{p},k} + \omega _ {\mathrm{s},-k}$ represents the energy of the magnons relevant to this process.
The expression \eqref{eq:vHs} is reminiscent of the van Hove singularity, which can be observed in optical absorption measurement, in that the quantity is inversely proportional to the derivative of the energy difference between the system and excitation.
In our case, the excitation is provided by the linear mechanical motion through the Doppler shift $vk$.
This perspective offers an additional understanding of the critical behaviour of the spin current.

Spin current as a function of the velocity of the probe magnon is shown in Figure\,\ref{fig2}(b). As we discussed, no spin current is generated below $v_\mathrm{cr}$. At $v=v_\mathrm{cr}$ the spin current has a peak, and the spin current becomes zero as the velocity increases. This trend qualitatively differs from the previous studies\,\cite{oue2022motion,oue2024optimizing}, which consider a spin-conserving tunneling Hamiltonian, where the spin current is quadratic with respect to velocity.

We would like to note that the spin current considered here is driven by the spin-nonconservative process at the interface $H_2$; thus, it is not from one magnet to the other. The interface, where the linear motion affects the magnonic system, can be a spin sink and source (i.e., it can annihilate magnons by $b _ {\mathrm{p},k} b _ {\mathrm{s},-k}$ and create magnons with $b _ {\mathrm{p},k} ^ \dagger b _ {\mathrm{s},-k} ^ \dagger$).
It is rather related to the conversion of the kinetic energy ($\propto \abs{v} ^ 2$) of the linear motion to the energy stored within the magnonic subsystem through the mechanism akin to parametric oscillation.
Therefore, $\abs{v}$ rather than $v$ plays a role, and the sample magnet always experiences spin current incoming from the interface, regardless of the direction of the linear motion, when the threshold is approached.

Let us consider the group velocities of magnons at $v_\mathrm{cr}$. We can interpret Hamiltonian $H_2$ as a creation(annihilation) of the sample `anti-magnon' $a_{\mathrm{s},k}^\dagger = b_{\mathrm{s},k}$($a_{\mathrm{s},k} = b_{\mathrm{s},k}^\dagger$) and an annihilation(creation) of the probe magnon $b_{\mathrm{p},-k}$($b_{\mathrm{p},-k}^\dagger$). At critical velocity $v_\mathrm{cr}$, we can calculate that the group velocities of the magnon and the anti-magnon coincide with each other,
\begin{align}
    \left.\dv{\overline{\omega}_{\mathrm{s},k}}{k}\right|_{v=v_\mathrm{cr}} 
    = \left.\dv{\qty(\omega_{\mathrm{p},k} - vk)}{k}\right|_{v=v_\mathrm{cr}},
\end{align}
where we defined anti-magnon dispersion as $\overline{\omega}_{\mathrm{s},k} \equiv -\omega_{\mathrm{s},-k}$.
Due to the group velocity matching, the two excitations run in parallel at identical velocities; hence, they engage in a strong mutual interaction at $v_\mathrm{cr}$, leading to the divergence of the spin current.

\begin{figure}
    \centering
    \includegraphics[width=\linewidth]{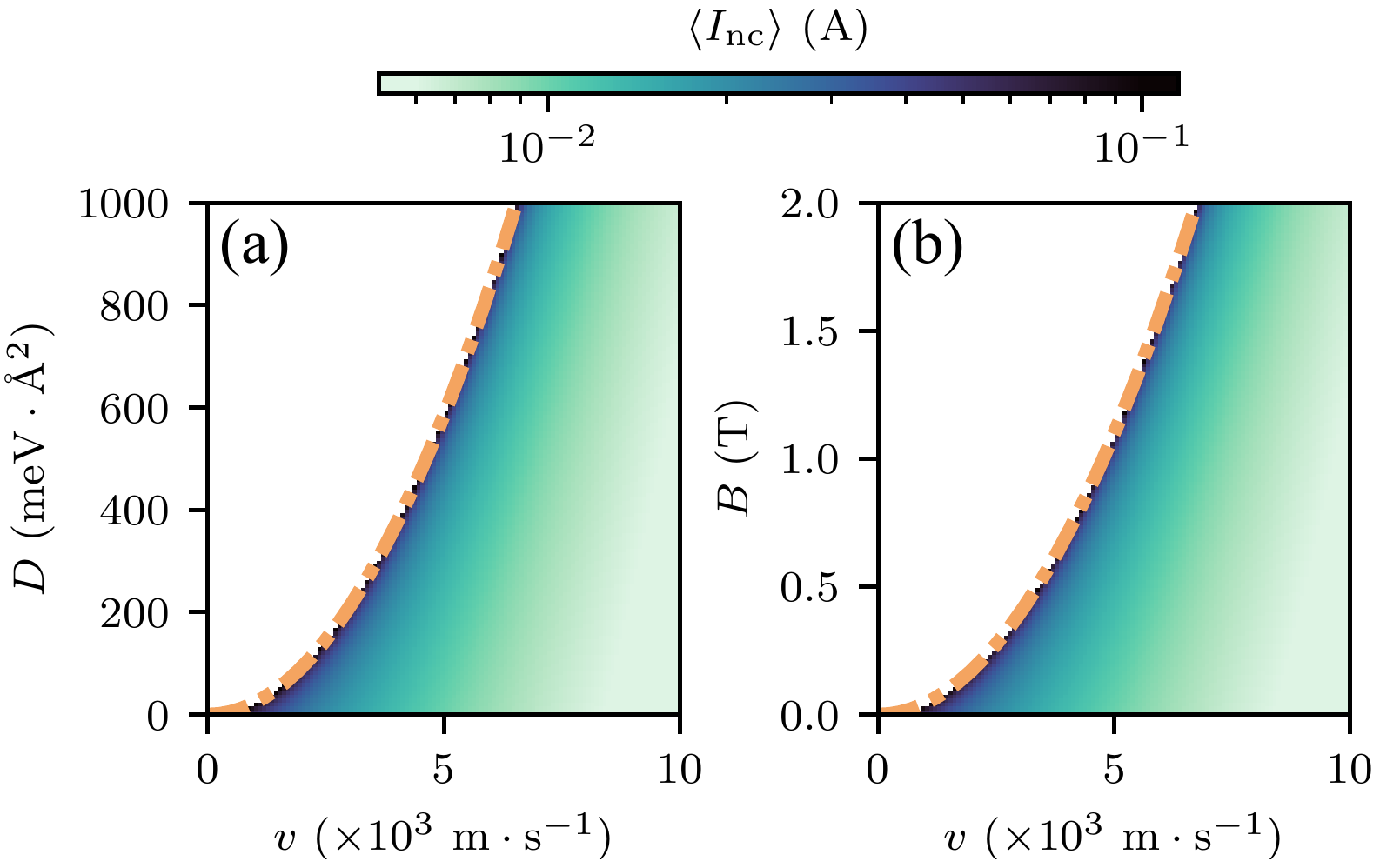}
    \caption{Spin current as functions of the velocity of probe medium $v$ and (a) $D$ (b) the static magnetic field $B$. We set $B=1~\rm(T)$ in (a), and $D=532~\rm(meV\cdot\text{\AA}^2)$, a value characteristic of yttrium iron garnet\,\cite{princep2017full} in (b). Dependences of $D$ and $B$ on the dispersion relation $\omega_k = Dk^2 +\omega_0~(\omega_0=\gamma B)$ cause the change of the amount of the spin current.  For spin currents to be produced, there must be crossing point(s) in Figure \ref{fig2}(a). From this condition, there is a critical velocity $v_\mathrm{cr} = 4\sqrt{D\omega_0}$ shown as the orange dashed line. The introduction of the effective mass of the magnon $D=\hbar/2m^*$ leads to the equivalent condition that the kinetic energy of the magnon $m^*v^2/2$ should be at least twice the ground state energy of the magnon $2\cdot2\hbar\omega_0$.}
    \label{fig3}
\end{figure}

Spin current as a function of the stiffness constant ($D$) and the magnetic field ($B$) with critical velocity $v_\mathrm{cr}$ is shown in Figure~\ref{fig3}. $v_\mathrm{cr}$ has a parabolic feature dependent both on $D$ and $B$. These trends arise from the fact that spin current generation is governed by the quadratic equation:\,$2Dk^2 -vk +2\omega_0 = 0$. In the case of yttrium iron garnet\,(YIG) with a spin-wave stiffness of $D=532\, \rm meV\cdot \AA^2$\,\cite{princep2017full}, the critical velocity for the spin current generation is $\sim 10^3\,\rm m\cdot s^{-1}$. To decrease critical velocity $v_{\rm cr}$, a lower-stiffness material can be adopted, such as amorphous ferromagnets\,\cite{grigoriev2022study} and artificial magnonic crystal structures with a flat band\,\cite{chen2022magic}.

\section{Discussions.}
Let us consider a unitary transformation which removes the Doppler shift in the dispersion relation~$\qty(\omega_{\mathrm{p},k} -vk \mapsto \omega_{\mathrm{p},k})$\,\cite{oue2024optimizing}: $U = \prod_k\exp \qty(-ivk\,b_{\mathrm{p},k}^\dagger b_{\mathrm{p},k}\,t)$.
With this unitary formation, the unperturbed $H _ 0$ and perturbed Hamiltonian $H_2$ are, respectively, transformed as follows: 
\begin{align}
    H _ 0 &\mapsto \sum_k \hbar\qty(\omega_{\mathrm{p},k} + vk)b_{\mathrm{p},k}^\dagger b_{\mathrm{p},k} + \sum_k\hbar\omega_{\mathrm{s},k}b_{\mathrm{s},k}^\dagger b_{\mathrm{s},k},
    \\
    H_2 &\mapsto \sum_k H_{\rm nc} \qty(e^{ivkt} b_{\mathrm{p},-k}b_{\mathrm{s},k} + e^{-ivkt} b_{\mathrm{p},-k}^\dagger b_{\mathrm{s},k}^\dagger).
    \label{eq: H _ 2'}
\end{align}
The interaction part~\eqref{eq: H _ 2'} in this representation bears a striking resemblance to the parametric down-conversion in nonlinear optics.
Within the semiclassical approximation, the Hamiltonian responsible for the parametric down-conversion is expressed as $H_{\rm int} \propto e^{-i\omega_e t} a_i^\dagger a_s^\dagger +\rm H.c.$\,\cite{scully1997quantum}, where \(\omega_e\) is the frequency of the excitation laser, and \(a_i^\dagger\) and \(a_s^\dagger\) are creation operators for the emitted photons (idler and signal, respectively).
In our scenario, the Doppler shift serves as the excitation akin to the excitation laser in parametric down-conversion, leading to a singularity when the matching condition is met.

Furthermore, our system is also closely related to the parallel pumping in spintronics and magnonics\,\cite{bracher2017parallel}, where a microwave pump field $\omega_p$ creates or annihilates magnon pairs.
The interaction Hamiltonian for parallel pumping $H_{\rm int}^{\rm PP}(\vb{k}) \propto e^{-i\omega_\mathrm{p} t} c^\dagger_{\vb{k}}c^\dagger_{-\vb{k}} + \mathrm{H.c.}$ bears a strong resemblance to our transformed Hamiltonian (\ref{eq: H _ 2'}), in which the sliding motion introduces a phase factor $e^{-ivkt}$ that effectively plays the role of an external excitation source, analogous to $e^{-i\omega_\mathrm{p} t}$.
While the parallel pumping generates magnon pairs in a single medium, we may have magnon pairs across two media.

In relation to that, we would also like to mention that the interaction~\eqref{eq: H _ 2'} is distinct from the linear excitation of magnons by an AC field.
Specifically, our interaction Hamiltonian $H _ 2$ describes two-mode squeezing, where the driving factor $e^{\pm ivkt}$ couples to two magnons.
In contrast, a conventional linear AC drive follows $H_{\rm drive} \propto e ^ {-i\Omega t} b ^ \dagger + e ^ {+i\Omega t} b$, where $\Omega$ is the driving Rabi frequency, coupling only to a single magnon mode.

As indicated by the phase factor $e^{\pm ivkt}$ in Eq.~\eqref{eq: H _ 2'}, the tunnelling interaction undergoes periodic modulation with the Doppler frequency $vk$.
If the sliding speed $v$ is excessively high, the temporal modulation would be too rapid, and the interaction would be effectively averaged out, allowing the application of the rotating wave approximation.
This is consistent with the observation that our current diminishes in the limit of large velocity in Figure\,\ref{fig2}(b).

\begin{figure}
    \centering
    \includegraphics[width=\linewidth]{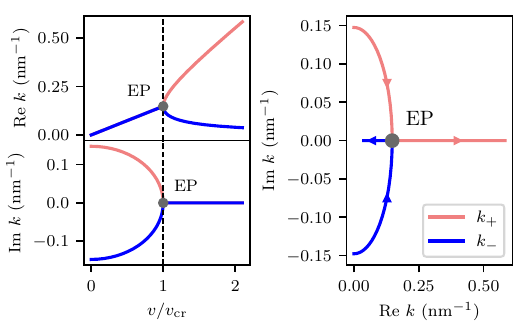}
    \caption{Parametric instability at a critical velocity $v_\mathrm{cr}$ where the real (imaginary) part of the solution (coalesces) and the spin current is resonantly enhanced.}
    \label{fig4}
\end{figure}

The real and imaginary part of the solutions of $\omega_{\mathrm{p},k} - vk + \omega_{\mathrm{s},-k} = 0$ as a function of probe magnon velocity is shown in Figure~\ref{fig4}. At critical velocity $v_\mathrm{cr}$, the real (imaginary) part splits~(coalesces). This behavior has a similarity to the exceptional point in photonics and acoustics\,\cite{ozdemir2019parity, miri2019exceptional, shi2016accessing}, which is the transition point between $PT$ symmetry and broken $PT$ symmetry. Indeed, we can formulate an eigenvalue problem based on the Heisenberg equations governing the annihilation and creation operators (See Appendix \ref{appx:PT-symmetry} for the detail), in which the eigenvalues undergo a transition from real values to complex values when crossing the critical point\,\cite{brevik2022fluctuational}. Furthermore, since only forward-propagating magnons ($k > 0$) are emitted and amplified, the proposed mechanism can be considered as unidirectional amplification, akin to lasing phenomena.
This phenomenon--where the motion of one magnet unidirectionally generates magnons in the other--could be analogous to magnon drag, wherein a magnon current in one magnet induces an effective magnonic potential in another magnet placed nearby, thereby generating a corresponding magnon current\,\cite{liu2016nonlocal}.
Generalizing our theory for finite temperature might provide a foundation for observing similar behaviors in the two different setups.

Finally, we emphasize the distinction from our previous studies\,\cite{oue2022motion, oue2024optimizing}, which considered a spin-conserving tunneling Hamiltonian at finite temperature.
While those works report a spin current that increases quadratically with the sliding velocity, our results reveal a critical behavior, characterized by a divergence of the spin current at a critical velocity.
At the low-temperature regime, the spin-conserving Hamiltonian yields no spin current, whereas the spin non-conserving Hamiltonian leads to a finite spin current originating from quantum fluctuations\,(see Appendix \ref{appx:derivation-spin-current}).

\section{Conclusion and outlook.}
We showed that a magnomechanical system consisting of two ferromagnetic insulators separated by a narrow gap, one of which moves at a constant velocity, can coherently and efficiently generate a spin current using parametric instability.
This technique offers an efficient method for generating spin currents and enables high-precision measurements of sliding velocity $v$, stiffness constant $D$, and magnetic field $B$ due to the distinct variation in spin current generation around the critical velocity $v_{\rm cr} \propto \sqrt{DB}$.

In addition to these practical implications, we identified and discussed several intriguing analogies and connections between our system and other areas of physics.
These include the Landau criterion for superfluidity, the van Hove singularities, parametric down-conversion in quantum optics, and spontaneous parity-time symmetry breaking. Such analogies not only deepen our understanding of the underlying mechanisms but also hint at broader interdisciplinary connections.
We expect that our results open a way to advance magnon-based quantum technology and quantum sensing.

\section*{Acknowledgement}
We are deeply grateful to Mamoru Matsuo for the fruitful discussions.
We also thank Shuichi Iwakiri, Yugo Onishi, Aruku Senoo, Hiroyuki Tajima, Keisuke Kato, Ken Hirata, and Yuri Nakamura for their comments.
T.U.~acknowledges support from the Funai Overseas Scholarship. D.O.~is supported by JSPS Overseas Research Fellowship, by the Institution of Engineering and Technology (IET), and by Funda\c{c}\~ao para a Ci\^encia e a Tecnologia and Instituto de Telecomunica\c{c}\~oes under project UIDB/50008/2020.

\appendix
\section{Interaction Hamiltonian}
\label{appx:derivation-Hint}
Let us assume the first (second) magnet occupies $y \geq L$ ($y \leq 0$) and denote the coordinate parallel to the surfaces by $\vb{x}$.
Consider the interaction between two magnetic dipoles, $\vb{p} _ \mathrm{m} ^ {(1)}$ and $\vb{p} _ \mathrm{m} ^ {(2)}$, which reside in the first and second magnets, respectively.
The magnetic field at $(\vb{x},y \geq L)$ generated by the second dipole $\vb{p} _ \mathrm{m} ^ {(2)}$ placed at $(\vb{x}',y' \leq 0)$ can be evaluated as $\vb{B}(\vb{x},y) = G(\vb{x},y|\vb{x}',y') \vb{p} _ \mathrm{m} ^ {(2)}(\vb{x}',y')$, where $G$ is Green's function derived from Maxwell's equations (The derivation is given in Appendix \ref{appx:greens-function}).
Thus, its influence on the first dipole $\vb{p} _ \mathrm{m} ^ {(1)}$ at $(\vb{x},z)$ is characterised by the interaction energy,
\begin{align}
    -\vb{p} _ \mathrm{m} ^ {(1)}(\vb{x},y) \cdot G(\vb{x},y|\vb{x}',y') \cdot \vb{p} _ \mathrm{m} ^ {(2)}(\vb{x}',y').
\end{align}
As the dipoles on the surfaces ($y = 0$ and $y = L$) should have the largest contributions, we shall focus on them by setting
$\vb{p} _ {\mathrm{m}} ^ {(1)}(\vb{x},y)\to \vb{p} _ {\mathrm{p}}(\vb{x})\delta(y-L)$
and
$\vb{p} _ {\mathrm{m}} ^ {(2)}(\vb{x}',y') \to \vb{p} _ {\mathrm{s}}(\vb{x}')\delta(y')$.
Performing the integration over $(\vb{x},y)$ and $(\vb{x}',y')$,
\begin{align}
    H _ \mathrm{int} = -\iint
    \vb{p} _ {\mathrm{p}}(\vb{x}) 
    \cdot G(\vb{x},L|\vb{x}',0) \cdot 
    \vb{p} _ {\mathrm{s}}(\vb{x}')
    \dd{\vb{x}}\dd{\vb{x}'}.
\end{align}

In the present case, the dipole vectors should be promoted to operators:
\begin{align}
    \vb{p} _ {\mathrm{p(s)}}(\vb{x})
    \to 
    \vb{p} _ {0}
    b _ {\mathrm{p(s)}}(\vb{x})
    +\vb{p} _ {0} ^ *
    b _ {\mathrm{p(s)}} ^ \dagger(\vb{x}),
    \label{eq:p_ps}
\end{align}
where we defined 
$
\vb{p} _ {0} =
-g\mu _ \mathrm{B}\sqrt{2S}\qty(
\vb{u} _ x - i\vb{u} _ y
),
$
and $b _ {\mathrm{p(s)}}(\vb{x})$ are the magnon operators acting on the first and second magnets.
Note that the $z$ component has been neglected in writing Eq.~\eqref{eq:p_ps}, as it does not contribute to the interaction at the lowest-order spin-wave approximation \footnote{
The $z$ component of the spin operator has the form
$
S _ {\mathrm{p,s}} ^ {z} = 
\hbar(S - b _ {\mathrm{p,s}} ^ \dagger b _ {\mathrm{p,s}}),
$
which is already at the second order in the magnon operator.
Thus, up to the lowest order in the spin-wave approximation (retaining the quadratic contributions only), we have
$
S _ {\mathrm{p}} ^ {z}S _ {\mathrm{s}} ^ {z} \approx
\hbar ^ 2 (S ^ 2 - S b _ {\mathrm{p}} ^ \dagger b _ {\mathrm{p}} - S b _ {\mathrm{s}} ^ \dagger b _ {\mathrm{s}}),
$
$
S _ {\mathrm{p}} ^ {z}S _ {\mathrm{s}} ^ {\pm} \approx
\hbar S S _ {\mathrm{s}} ^ {\pm},
$
and
$
S _ {\mathrm{p}} ^ {\pm}S _ {\mathrm{s}} ^ {z} \approx
\hbar S S _ {\mathrm{p}} ^ {\pm},
$.
These do not contribute to the interaction.
}.
The magnon operator can be expanded as $b _ {\mathrm{p(s)}}(\vb{x}) = \sum _ {\vb{k}} b _ {\mathrm{p(s)},\vb{k}}e ^ {i\vb{k}\cdot\vb{x}}$.
Substituting, we can obtain the effective interaction Hamiltonian,
\begin{align}
    H _ \mathrm{int} 
    = &\sum _ {\vb{k},\vb{k}'} \qty{
    H _ \mathrm{nc}(\vb{k},\vb{k}') b _ {\mathrm{p},\vb{k}} b _ {\mathrm{s},\vb{k}'}
    +H _ \mathrm{nc} ^ *(\vb{k},\vb{k}') b _ {\mathrm{p},\vb{k}} ^ \dagger b _ {\mathrm{s},\vb{k}'} ^ \dagger
    } \notag \\
    + &\sum _ {\vb{k},\vb{k}'} \qty{
    H _ \mathrm{ex}(\vb{k},\vb{k}') b _ {\mathrm{p},\vb{k}} ^ \dagger b _ {\mathrm{s},\vb{k}'}
    +H _ \mathrm{ex} ^ *(\vb{k},\vb{k}') b _ {\mathrm{p},\vb{k}} b _ {\mathrm{s},\vb{k}'} ^ \dagger
    },
    \label{eq:Hint}
\end{align}
where we defined the coupling constants,
\begin{align}
    &H _ \mathrm{nc}(\vb{k},\vb{k}') = \iint
    \vb{p} _ {0}
    e ^ {i\vb{k}\cdot\vb{x}}
    \cdot G(\vb{x}|\vb{x}') \cdot
    \vb{p} _ {0}
    e ^ {i\vb{k}'\cdot\vb{x}'}
    \dd{\vb{x}}\dd{\vb{x}'},
    \\
    &H _ \mathrm{ex}(\vb{k},\vb{k}') = \iint 
    \vb{p} _ {0} ^ * e ^ {-i\vb{k}\cdot\vb{x}}
    \cdot G(\vb{x}|\vb{x}') \cdot
    \vb{p} _ {0} e ^ {i\vb{k}'\cdot\vb{x}'}
    \dd{\vb{x}}\dd{\vb{x}'}.
\end{align}
Note that we suppressed the second argument for conciseness [i.e., $G(\vb{x}|\vb{x}') := G(\vb{x},L|\vb{x}',0)$].
Here, we substitute the free-space magnetostatic Green's function, 
$
    G(\vb{x}|\vb{x}') \to G _ 0(\vb{x}|\vb{x}') 
    = \int 
    G _ {\vb{k}} ^ {(0)} e ^ {i\vb{k}\cdot(\vb{x}-\vb{x}')}
    \dd{\vb{k}}.
$
(To take into account the effects of inhomogeneity and bulk contributions, we should modify the Green's function.)
Then, the coupling channel will be restricted to the momentum-conserving one,
\begin{align}
    &H _ \mathrm{nc}(\vb{k},\vb{k}') \to 
    \vb{p} _ {0} \cdot G _ {-\vb{k}} ^ {(0)} \cdot \vb{p} _ {0} \delta(\vb{k}+\vb{k}')
    =: H _ \mathrm{nc} \delta(\vb{k}+\vb{k}'),
    \\
    &H _ \mathrm{ex}(\vb{k},\vb{k}') \to
    \vb{p} _ {0} ^ * \cdot G _ {\vb{k}} ^ {(0)} \cdot \vb{p} _ {0}
    \delta(\vb{k}-\vb{k}')
    =: H _ \mathrm{ex} \delta(\vb{k}-\vb{k}').
\end{align}
Note that both of the interaction coefficients are real-valued, $H _ \mathrm{nc(ex)} ^ * = H _ \mathrm{nc(ex)}$.
Overall, the interaction Hamiltonian is approximately given as
\begin{align}
    H _ \mathrm{int} \to 
    &\sum _ {\vb{k}} \qty{
    H _ \mathrm{nc} b _ {\mathrm{p},\vb{k}} b _ {\mathrm{s},-\vb{k}}
    +H _ \mathrm{nc} b _ {\mathrm{p},\vb{k}} ^ \dagger b _ {\mathrm{s},-\vb{k}} ^ \dagger
    }
    \\ \notag
    &+ \sum _ {\vb{k}} \qty{
    H _ \mathrm{ex} b _ {\mathrm{p},\vb{k}} ^ \dagger b _ {\mathrm{s},\vb{k}}
    +H _ \mathrm{ex} b _ {\mathrm{p},\vb{k}} b _ {\mathrm{s},\vb{k}} ^ \dagger
    }.
    \label{eq:Hint_approx}
\end{align}
Since the magnon modes running parallel or anti-parallel to the mechanical linear motion are affected most, we focused on the direction of the linear motion in the manuscript.

\section{Magnetostatic Green's function}
\label{appx:greens-function}
The magnetostatic Green's function in free space can be found as follows:
In the non-relativistic limit, Maxwell's equations give
\begin{align}
    \nabla \cdot \vb{H} = -\nabla \cdot \vb{M},
    \quad
    \nabla \times \vb{H} = 0.
\end{align}
The second equation implies that we can write the magnetic field in terms of `the magnetic scalar potential,' $\vb{H} = -\nabla \psi$.
Substituting this into the first equation, we can get
$-\nabla ^ 2 \psi = -\nabla \cdot \vb{M}.$
Assuming $\vb{M} = \vb{p} _ \mathrm{m} \delta(\vb{r} - \vb{r}')$ to evaluate the Green's function, we can write
\begin{align}
    -\nabla ^ 2 \psi(\vb{r}) = -\nabla \cdot \vb{p} _ \mathrm{m}\delta(\vb{r}-\vb{r}').
\end{align}
This is nothing but the Poisson equation.
Using the Green's function for the Laplace equation $g _ 0(\vb{r}|\vb{r}')$, the magnetic scalar potential may be written as follows:
\begin{align}
    \psi(\vb{r}) 
    &= -\int 
    g _ 0(\vb{r}|\vb{r}'')
    \nabla'' \cdot \vb{p} _ \mathrm{m} \delta(\vb{r}'' - \vb{r}')
    \dd{\vb{r}''},
    \\
    &=\int \qty{\nabla'' g _ 0(\vb{r}|\vb{r}'')} \cdot \vb{p} _ \mathrm{m} \delta(\vb{r}'' - \vb{r}')\dd{\vb{r}'}.
\end{align}
Hence, our magnetic field will be
\begin{align}
    \vb{H}(\vb{r}) 
    &= 
    \int 
    \qty{-\nabla\nabla'' g _ 0(\vb{r}|\vb{r}'')} \cdot \vb{p} _ \mathrm{m} \delta(\vb{r}'' - \vb{r}')\dd{\vb{r}'}.
\end{align}
The quantity in the curly brackets gives our Green's function, $G _ 0(\vb{r}|\vb{r}')=-\mu _ 0\nabla\nabla' g _ 0(\vb{r}|\vb{r}').$
We use the Fourier representation to write down Green's function for the Laplace equation as follows:
\begin{align}
    g _ 0(\vb{r}|\vb{r}') 
    &= \int 
    \frac{1}{k _ y ^ 2 + \abs{\vb{k}} ^ 2}
    e ^ {i\vb{k}\cdot(\vb{x}-\vb{x}')+ik _ y(y-y')}
    \dd{k _ y}\dd{\vb{k}},
    \\ \notag
    &=
    \int 
    \frac{1}{2\abs{\vb{k}}}
    e ^ {i\vb{k}\cdot(\vb{x}-\vb{x}')-\abs{\vb{k}}\abs{y-y'}}
    \dd{\vb{k}},
\end{align}
where we defined $\dd{\vb{k}} := \dd{k _ x}\dd{k _ z}/(2\pi) ^ 2$.
Note that the integration over $k _ y$ has been performed on the complex plane.
Therefore, for $y > y'$, we can explicitly write
\begin{align}
    G _ 0(\vb{x},y|\vb{x}',y') 
    &=
    -\mu _ 0\int
    \frac{\vb{k} _ +\vb{k} _ +}{2\abs{\vb{k}}}
    e ^ {-\abs{\vb{k}}(y-y')}
    e ^ {i\vb{k}\cdot(\vb{x}-\vb{x}')}
    \dd{\vb{k}}
\end{align}
where we defined $\vb{k} _ + = \vb{k} + i\abs{k}\vb{u} _ y$.

\section{Derivation of the spin current formula}
\label{appx:derivation-spin-current}
We consider two ferromagnetic insulators separated by a narrow gap. 
Within the Holstein-Primakoff approximation, each magnet hosts magnon described by the following unperturbed Hamiltonian:
\begin{align}
    H_0 = \sum_{k\eta} \hbar\omega_{\eta,k} b_{\eta,k}^\dagger b_{\eta,k},
\end{align}
where $\eta=\mathrm{p}, \mathrm{s}$ specify the magnet, and $\omega_{\eta, k}$ is the magnon dispersion, and $b_{\eta,k}^{(\dagger)}$ annihilates$\,$(creates) magnons in the magnet $\eta$. 

The interaction between the two magnets involves two contributions $H_{\rm int} = H_1 + H_2$ with
\begin{align}
    H_1 &= \sum_k H_{\rm ex}b_{\mathrm{s},k}^\dagger b_{\mathrm{p},k} + \rm H.c., \nonumber \\
    H_2 &= \sum_k H_{\rm nc}b_{\mathrm{s},k} b_{\mathrm{p},-k} + \rm H.c. \nonumber,
\end{align}
where $H_{\rm ex}$ and $H_{\rm nc}$ are coupling strengths. Here, for simplicity, we consider the constant coupling strengths.

\subsection{Spin current between two magnets.}
In the interaction picture, the change in total spin within the sample medium can be obtained as
\begin{align}
    \pdv{t}\sum_k \expval{S_{\mathrm{s},k}^z\qty(t)} &= -\sum_k 2H_{\rm ex} \mathrm{Im}\expval{b_{\mathrm{s},k}^\dagger\qty(t)b_{\mathrm{p},k}\qty(t)} \nonumber\\
    &~~~+\sum_k 2H_{\rm nc} \mathrm{Im}\expval{b_{\mathrm{s},-k}\qty(t)b_{\mathrm{p},k}\qty(t) }.
\end{align}
where $S_{\mathrm{s},k}^z = S - b_{\mathrm{s},k}^\dagger b_{\mathrm{s},k}$ is the $z$ component of the spin in the sample medium, and $\expval{\cdot}$ denotes the average with respect to the full Hamiltonian. We define the spin current flowing into the sample medium at $t=t_1$ as
\begin{align}
    \expval{I_s(t_1)} &\equiv \expval{I_{\rm ex}(t_1)} + \expval{I_{\rm nc}(t_1)} \nonumber\\
    &= \sum_k 2H_{\rm ex} \mathrm{Im}\expval{\mathcal{T}\;b_{\mathrm{p},k}(t_1-0)b_{\mathrm{s},k}^\dagger(t_1)} \nonumber\\
    &~~~- \sum_k 2H_{\rm nc} \mathrm{Im}\expval{\mathcal{T}\;b_{\mathrm{p},k}(t_1-0)b_{\mathrm{s},-k}(t_1)},
\end{align}
where $\mathcal{T}$ is the time-ordering operator.
Let us consider the low-temperature ($T\rightarrow0$) and lossless limit. The first term $\expval{I_{\rm ex}(t_1)}$ does not contribute to the spin current in this limit due to the factor related to the distribution function. On the other hand, $\expval{I_{\rm nc}(t_1)}$ remains in this limit. This means $\expval{I_{\rm nc}(t_1)}$ is purely originated from quantum fluctuation. Let us focus on this term  $\expval{I_{\rm nc}(t_1)}$. Using the perturbative expansion, we can evaluate the spin current up to the second order in the coupling strength $H_{\rm nc}$,
\begin{align}
    \expval{I_{\rm nc} (t_1)} 
    &= \frac{2H_{\rm nc}^2}{\hbar}\sum_{k}\mathrm{Re}\int_C \dd t_2 \expval{\mathcal{T}_C b_{\mathrm{p},k}(t_1^+))b_{\mathrm{p},k}^\dagger(t_2)}_0 \nonumber\\
    &~~~~~~~~~~~~~~~~\times\expval{\mathcal{T}_Cb_{\mathrm{s},-k}(t_1^-)b_{\mathrm{s},-k}^\dagger(t_2)}_0,
\end{align}
where $\expval{\cdot}_0$ is the average with respect to the unperturbed Hamiltonian\,$H_0$, and $\mathcal{T}_c$ is the time-ordering operator on the Schwinger-Keldysh contour composed of a forward branch $C_+$ and a backward one $C_-$. Note that $t^\pm$ denotes time on the forward and backward branches.

Here, we introduce non-equilibrium Green's function,
\begin{align}
    \chi(t_1, t_2) \equiv \frac{1}{i\hbar}\expval{\mathcal{T}_C b_k(t_1)b_k^\dagger(t_2)}_0,
\end{align}
whose lesser and greater components read
\begin{align*}
    \chi_{k;12}^< \equiv \chi_k(t_1^+,t_2^-) = \frac{1}{i\hbar}\expval{b_k^\dagger(t_2)b_k(t_1)}_0 \\
    \chi_{k;12}^> \equiv \chi_k(t_1^-,t_2^+) = \frac{1}{i\hbar}\expval{b_k(t_1)b_k^\dagger(t_2)}_0.
\end{align*}
We can write the chronologically ordered and anti-chronologically ordered components in terms of the lesser and greater components,
\begin{align*}
    \chi_{k;12}^{++} \equiv \theta(t_1-t_2)\chi_{k;12}^> + \theta(t_2-t_1)\chi_{k;12}^<, \\
    \chi_{k;12}^{--} \equiv \theta(t_1-t_2)\chi_{k;12}^< + \theta(t_2-t_1)\chi_{k;12}^>,
\end{align*}
where $\theta$ denotes the Heaviside unit step function.
\begin{align}\label{eq: time domain}
    \frac{\expval{I_{\rm nc}(t_1)}}{2\hbar H_{\rm nc}^2} = -\sum_k \mathrm{Re} \int_{-\infty}^{\infty}\dd t_2 \chi_{\mathrm{p},k;12}^\mathfrak{R}\chi_{\mathrm{s},-k;12}^>,
\end{align}
where we have defined the retarded component,
\begin{align}
    \chi_{k;12}^\mathfrak{R} &\equiv \frac{1}{i\hbar}\theta(t_1-t_2)\expval{\comm{b_k(t_1)}{b_k^\dagger(t_2)}}_0 \nonumber\\
    &= \chi_{k;12}^{++} - \chi_{k;12}^<,
\end{align}
which is nothing but the dynamical (magnetic) susceptibility of the magnet.

In the steady state, Green's functions depend only on the time difference $t_{12} = t_1 - t_2$, e.g.,
\begin{align}
    \chi_{k;12}^\mathfrak{R} = \frac{1}{2\pi}\int\dd\omega\chi_{k\omega}^\mathfrak{R}e^{-i\omega t_{12}}.
\end{align}
Thus, working on the frequency domain, we can simplify the convolution-type integral \eqref{eq: time domain},
\begin{align}\label{eq: current_ap}
    \expval{I_{\rm nc}(t)} = \frac{8\pi H_{\rm nc}^2}{\hbar}\sum_k \delta(\omega_{\mathrm{p},k} + \omega_{\mathrm{s},-k}).
\end{align}

\section{$PT$ symmetry}
\label{appx:PT-symmetry}
Our system is inherently out of equilibrium because of the non-uniform motion (shearing motion) and cannot be mapped to a system at rest by the global Galilean transformation.
Thus, the effective description of the system may be provided using the non-Hermitian matrix (in our formulation, we will use the Liouvillian matrix).
It is evident that parity symmetry is broken due to the non-uniform motion, and time-reversal symmetry is also broken due to the linear motion.
On the other hand, with the sequential application of the parity and time-reversal transformation, we can retain the configuration invariant.
Therefore, our nonequilibrium system can be characterised by the $PT$ symmetry.

In particular, here, we show that the threshold for the spin current generation corresponds to the $PT$-symmetry breaking (exceptional) point.
Let us consider the Heisenberg equations for one-body operators.
We can write them in a matrix from:
\begin{align}
    i\pdv{t}\mqty(b_{\mathrm{s},-k}\\ b_{\mathrm{p},k}\\ b_{\mathrm{s},-k}^\dagger\\ b_{\mathrm{p},k}^\dagger) = \mqty(\omega_{\mathrm{s},-k}& 0& 0& g\\
    0& \omega_{\mathrm{p},k}& g& 0\\
    0& -g& -\omega_{\mathrm{s},-k}& 0\\
    -g& 0& 0& -\omega_{\mathrm{p},k}) 
    \,\mqty(b_{\mathrm{s},-k}\\ b_{\mathrm{p},k}\\ b_{\mathrm{s},-k}^\dagger\\ b_{\mathrm{p},k}^\dagger),
    \label{eq: Heisenberg}
\end{align}
where we defined $g \equiv H_{\rm nc}/\hbar$ for convenience.
This equation is compatible with the coupled-mode theory~\cite{haus1991coupled}, which is widely used in studying $PT$-symmetry-related phenomena in optics~\cite{zyablovsky2014pt,zhao2018parity,ozdemir2019parity,krasnok2019anomalies}.
The coupled-mode theory gives the time evolution of modal amplitudes, which may be promoted to bosonic operators; thus, it is compatible with the Heisenberg equations for the one-body operators.
In other words, we may take the expectation value of the bosonic operators with respect to a coherent state to effectively substitute classical amplitudes into the operators.

The time evolution of the bosonic operators given in Eq.~\eqref{eq: Heisenberg} is controlled by the Liouvillian matrix,
\begin{align}
    \mathcal{L} = \mqty(\omega_{\mathrm{s},-k}& 0& 0& g\\
    0& \omega_{\mathrm{p},k}& g& 0\\
    0& -g& -\omega_{\mathrm{s},-k}& 0\\
    -g& 0& 0& -\omega_{\mathrm{p},k}), \label{eq: Liouvillian}
\end{align}

To establish the connection to the $PT$ symmetry, it is convenient to work in a ``centre-of-mass'' frame, where two ferromagnets move in the opposite directions at a speed of $v/2$ so that the Doppler shift is applied to each magnet in an opposite way (i.e.~one can write $\omega _ {\mathrm{s},k} = Dk ^ 2 + \omega _ 0 + kv/2$ and $\omega _ {\mathrm{p},k} = Dk ^ 2 + \omega _ 0 - kv/2$).
The time-reversal operation flips the sign of all time-related quantities (for the present case, $v \mapsto -v$), whereas the parity-reversal operation alters the sign of all coordinates so that we shall swap the labels,$(\mathrm{s},\mathrm{p}) \mapsto (\mathrm{p},\mathrm{s})$ and flip the sign of the wavenumber, $k \mapsto -k$.
The Liouvillian\,$\mathcal{L}$ is invariant under both the parity- and time-reversal operations; hence, our equation system may manifest two distinct regimes: $PT$-symmetry unbroken and broken phases.
The transition point between the two phases is called the exceptional point~\cite{zhao2018parity,ozdemir2019parity,krasnok2019anomalies}.
In the $PT$-symmetry unbroken (broken) phase, the eigenvalues of the matrix are real-valued (complex-valued)~\cite{zyablovsky2014pt}.

Indeed, the eigenvalues of the Liouvillian matrix~\eqref{eq: Liouvillian} can readily be found as
\begin{align}
    \pm\frac{1}{2}\qty{(\omega_{\mathrm{s},-k} - \omega_{\mathrm{p},k}) \pm \sqrt{(\omega_{\mathrm{s},-k} + \omega_{\mathrm{p},k})^2 -4g^2}},
    \label{eq:eigenvalues}
\end{align} 

\begin{figure*}[htbp]
    \centering
    \includegraphics[width=\linewidth]{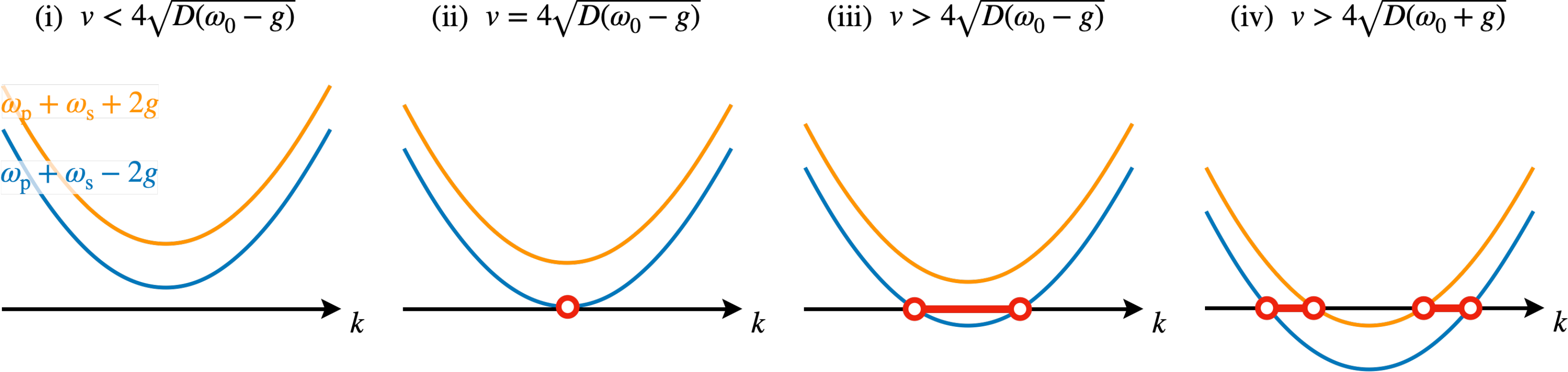}
    \caption{Two factors in the square root for various velocities $v > 0$. The square root in Eq.~\eqref{eq:eigenvalues} can be imaginary for sufficiently large velocities [for wavenumbers on the red line(s) in (iii) and (iv)].}
    \label{fig:ep}
\end{figure*}

which are either real-valued or complex-valued.
In order to find the eigenvalues are either real or complex numbers, we shall analyse the expression in the square root.
It can be factorised as
\begin{align}
    (2D k^2 - vk + 2\omega _ 0 - 2 g)(2D k^2 - vk + 2\omega _ 0 + 2 g).
    \label{eq:sqrt^2}
\end{align}
Being mindful of the fact that one of the two factors is always larger than the other,
\begin{align}
    2D k^2 - vk + 2\omega _ 0 + 2 g > 2D k^2 - vk + 2\omega _ 0 - 2 g,
\end{align}
we can find that the expression \eqref{eq:sqrt^2} yields a negative value for, at least, one wavenumber $k$, if one has a sufficiently large velocity,
\begin{align}
    v > 4\sqrt{D(\omega_0 - g)}.
\end{align}
Otherwise, both of the factors \eqref{eq:sqrt^2} are positive for all $k$, and the square root always returns a real number (see FIG.~\ref{fig:ep} for the visualisation).
The exceptional point is given by $v = 4\sqrt{D(\omega_0 - g)}$.

In the weak interaction regime ($g \ll \omega _ 0$), we can approximate the right-hand side of the $4\sqrt{D\omega_0}$, which is nothing but the threshold given by the perturbative analysis [Eq.~(3) in the main text].
Beyond this critical velocity, one of the eigenvalues of the Liouvillian acquires a positive imaginary part; hence, the time evolution of the operators becomes non-unitary (i.e.~the $PT$ symmetry is broken, and the corresponding eigenmode exponentially grows in time).
The spin current generation can be viewed as an outcome of the $PT$ symmetry breaking.

It may also be worth noting that the Liouvillian matrix~\eqref{eq: Liouvillian} can be decomposed as
\begin{align}
    \mathcal{L} \to \mqty(
    1& 0\\
    0& -1
    )
    \otimes
    \mqty(
    \omega_{\mathrm{s},-k}& g\\
    -g& -\omega_{\mathrm{p},k}
    ).
\end{align}
The matrix in the second entry in the tensor decomposition can be further unitary-transformed to a typical matrix that is studied in non-Hermitian physics in the context of $PT$ symmetry (e.g., systems with balanced loss and gain) \cite{el2018non}:
\begin{align}
    &\frac{1}{\sqrt{2}}
    \mqty(
    1& -i\\
    1& +i
    )
    \mqty(
    \omega_{\mathrm{s},-k}& g\\
    -g& -\omega_{\mathrm{p},k}
    )
    \frac{1}{\sqrt{2}}
    \mqty(
    1& 1\\
    i& -i
    )
    \notag \\
    &=\mqty(
    +ig& -kv/2\\
    -kv/2& -ig
    ),
\end{align}
It is clear that we have two modes experiencing balanced gain ($+ig$) and loss ($-ig$), which stem from the non-conserving part of the interaction Hamiltonian. The two modes are mutually coupled through the Doppler effect $-kv/2$.

\bibliographystyle{apsrev4-1}
\bibliography{reference}

\end{document}